\documentclass[12pt]{article}

\usepackage{color}
\usepackage{amsmath}
\usepackage{mathrsfs}
\usepackage{a4wide}
\usepackage{epsfig}
\usepackage{theorem}
\usepackage{amssymb}
\usepackage{bbm}

\usepackage[T1]{fontenc}
\usepackage[latin1]{inputenc}

\newcommand{\cN}{{\cal N}}

\newcommand{\bi}{\bigskip}
\newcommand{\no}{\noindent}
\newcommand{\be}{\begin{eqnarray}}
\newcommand{\ee}{\end{eqnarray}}
\newcommand{\ba}{\begin{align}}
\newcommand{\ea}{\end{align}}

\newcommand{\hk}{\hspace{0.1cm}}

\newcommand{\rk}{\right)}
\newcommand{\lk}{\left(}

\newcommand{\MM}{\mathsf{M}}

\newcommand{\vek}[1]{\mathbf{#1}}

\begin{document}

\title{The role of center vortices in Gribov's confinement scenario}

\author{M. Quandt, H. Reinhardt, G. Burgio \\Institut f\"ur Theoretische Physik\\
Auf der Morgenstelle 14\\
D-72076 T\"ubingen\\
Germany}
\date{\today}

\bi

\no

\maketitle
\bi

\no
\begin{abstract}
The connection of Gribov's confinement scenario in Coulomb gauge with the center
vortex picture  of confinement is investigated. For
this purpose we assume a vacuum wave functional which models the
infrared properties of the theory and in particular shows
strict confinement, i.e. an
area law of the Wilson loop. We isolate the center vortex content of this wave
functional by standard lattice methods and investigate their contributions to
various static propagators of the Hamilton approach to Yang-Mills theory in
Coulomb gauge. We find that the infrared properties of these quantities, in
particular the infrared divergence of the ghost form factor, are dominated by
center vortices.
\end{abstract}
\bi

\no
\section{Introduction}
\bi

\no
The infrared sector of QCD, in particular the confinement mechanism, is not
fully understood yet, although substantial progress had been made during recent
years. The progress comes mainly from lattice calculations \cite{R8},
\cite{RX1}, which gave support
to both the dual Meissner effect \cite{Nambu:1974zg}
and the center vortex picture \cite{'t:1977hy} of confinement
and also indicate that these two pictures are likely only two sides of the same
medal. Furthermore, there is lattice evidence that also Gribov's confinement
scenario is triggered by magnetic monopoles and center vortex
configurations \cite{Greensite:2003xf,Greensite:2004ur}.
In addition, one can show analytically that Zwanziger's horizon condition, i.e.
an infrared diverging ghost form factor, which is at the heart of Gribov's
confinement scenario in Coulomb gauge, implies dual Meissner effect
\cite{Reinhardt:2008ek}.
\bi

\no
In recent years there has been a renewed interest in studying Yang-Mills
theory in
Coulomb gauge, both in the continuum \cite{Zwanziger:2002sh} and on the
lattice \cite{Langfeld:2004qs, Quandt:2007qd, Voigt:2007wd, Voigt:2008rr,%
Burgio:2008jr,Burgio:2009xp}.
In particular, much
work was devoted to a variational solution of the Yang-Mills Schr\"odinger
equation in Coulomb gauge
\cite{Szczepaniak:2001rg,Feuchter:2004mk,Feuchter:2004gb,Reinhardt:2004mm,%
Epple:2006hv,Epple:2007ut,Reinhardt:2008ij,Schleifenbaum:2006bq,%
Reinhardt:2007wh}.
Using Gaussian type of Ansatzes for the vacuum wave functional, a set of
Dyson-Schwinger equations for the gluon and ghost propagators was derived
by minimizing the vacuum energy density. An infrared analysis
\cite{Schleifenbaum:2006bq}
of these equations exhibits solutions in accord with Gribov(-Zwanziger)
confinement scenario. Imposing Zwanziger's horizon condition one finds an
infrared diverging gluon energy and a linear rising static quark (Coulomb)
potential \cite{Schleifenbaum:2006bq} and also a perimeter law for the 't Hooft
loop \cite{Reinhardt:2007wh}. These infrared properties are reproduced by a
full
numerical solution of the Dyson-Schwinger equations over the entire momentum
regime \cite{Epple:2006hv}, and are also supported by lattice calculations
\cite{Quandt:2007qd,Burgio:2008jr,Burgio:2009xp}. Moreover, these lattice calculations 
also show clear
evidence for an infrared divergent ghost form factor and an infrared
suppressed static gluon propagator\footnote{This is different from Landau
gauge,
where lattice calculations seem to indicate an infrared finite gluon
propagator and ghost
dressing function
\cite{decoupling}, in contradiction to the scaling solution of the
Dyson-Schwinger
equations \cite{scaling}.}.
While the horizon condition has to be imposed \emph{by hand} in $D=3+1$,
as there exist also subcritical solutions with an infrared
finite ghost form factor\footnote{These subcritical solutions are the
analogue of the so-called de-coupling solution of the Dyson-Schwinger equations
in Landau gauge.} \cite{Epple:2007ut},
the coupled set of Dyson-Schwinger equations in $D = 2 + 1$  only allows
for critical solutions having an infrared diverging ghost form factor
\cite{Feuchter:2007mq}.
Furthermore, recent studies within the
functional renormalization group treatment of the Hamilton approach in Coulomb
gauge yields the horizon condition as a solution of the flow equation \cite{R19}.
In $D = 1+1$, finally, the exact ghost form factor is infrared enhanced
\cite{Reinhardt:2008ij}\footnote{Yang-Mills theory in $D = 1 + 1$ is
non-trivial
only on a compact manifold. In the Hamiltonian approach (with a continuous
time) space is $S_1$ so that the momenta are discrete. The zero momentum
corresponds to a zero mode of the Faddeev-Popov kernel and is therefore
excluded \cite{Reinhardt:2008ij}.}.
\bi

\no
Despite the encouraging results for the infrared properties of the various
Greens function (in particular the linear Coulomb potential) and the good
agreement of the gluon energy \cite{Epple:2007ut} with the lattice data
\cite{Burgio:2008jr,Burgio:2009xp}, the crucial test for the wave functional in the
variational approach, namely the calculation of the Wilson
loop, still has to come. In fact, a linear Coulomb potential is necessary but 
not sufficient for confinement
since the Coulomb string tension is only an upper bound to the Wilsonian string
tension \cite{Zwanziger:2002sh}. With the variational vacuum wave functional
at hand it would be, in principle,
straightforward to calculate the Wilson loop. However, path ordering makes
an exact evaluation of the Wilson loop impossible. In Ref.~\cite{Pak:2009em}
the spatial Wilson loop was calculated from a Dyson equation, which takes
care of the path ordering in an approximate fashion, employing the static
gluon propagator as input. Despite the rather limited range of
applicability of the Dyson equation, a linearly rising potential could be
extracted from the obtained Wilson loop.

Here we will proceed along a different line. We assume a wave functional
which is known \emph{a priori} to produce an area law for the (spatial) Wilson
loop, and calculate with this wave functional the various propagators
of the Hamiltonian approach in Coulomb gauge. The infrared properties of
these propagators are then compared with the ones obtained from
lattice results in $D=4$ Coulomb gauge which, in turn, agree qualitatively
with findings from a Gaussian type of variational wave functional.
\bi

\no
A simple choice for a confining wave functional is
\be
\label{1}
\Psi [\vek{A}] = \cN \exp \lk - \frac{1}{8 \mu}
\int d^3 x F_{i j} (\vek{x}) \,F_{i j} (\vek{x})\rk \hk ,
\ee
where $F_{ij}$ denotes the spatial components of the non-Abelian
field strength and $\mu$ is a dimensionfull parameter, which, in principle,
could serve as variational parameter. We will later see that the scaling
properties of $D=3$ Yang-Mills theory tie the parameter $\mu$ to the
numerical value of the (spatial) string tension: $\mu$ thus merely sets
the overall scale.
\bi

\no
The wave functional in eq.~(\ref{1}) models the infrared sector of the Yang-Mills
vacuum: it is gauge invariant and can be considered as the leading order in a
gradient expansion of the true Yang-Mills vacuum wave
functional \cite{R20}.
Furthermore, the functional of eq.~(\ref{1}) produces an area law for the (spatial)
Wilson loop. This is because the $D=4$ expectation value of any gauge invariant
and $A_0$--independent observable $\Omega[\vek{A}]$ in this state is precisely
given by the one in the $D = 3$ dimensional Yang-Mills theory,
\be
\label{2}
\langle \,\Psi\,|\,\Omega\,|\,\Psi \rangle =
\frac{\displaystyle\int \mathscr{D}\vek{A}\,\,
\Omega[\vek{A}]\,\exp\left[
\displaystyle -\frac{1}{4 \mu} \int d^3 x F^2_{ij}(\vek{x})\right]}
{\displaystyle\int\mathscr{D}\vek{A}\,
\exp\left[- \displaystyle\frac{1}{4\mu}
\int d^3 x F^2_{i j}(\vek{x})\right]}\,.
\ee
For gauge \emph{variant} observables such as the Green functions, we need to
pick a specific gauge on the vector potential $\vek{A}$. Choosing
Coulomb gauge $\nabla\,\vek{A} = 0$ in $D=4$ obviously entails
Landau gauge for the $D=3$ YM-theory.
\bi

\no
The wave functional of eq.~(\ref{1}) is certainly inappropriate
at large momenta where it yields a gluon energy
$\omega (|\vek{k}|) \sim |\vek{k}|^2$ instead of $\omega(\vek{k}) \sim
| \vek{k} |$. However, this
should be irrelevant for the confining properties. Also in the deep IR
region one does not expect eq.~(\ref{1}) to exactly reproduce the $D=4$
Yang-Mills Theory, since the standard lattice gluon and ghost propagator 
in $D=3$ Landau gauge
rather satisfy a decoupling type of solution \cite{decoupling},
in contrast to $D=4$ Coulomb gauge \cite{Burgio:2008jr,Burgio:2009xp}. Indeed,
one would expect the correct wave-function to be better described by
\cite{Greensite:2007ij}:
\be
\Psi [\vek{A}] = \cN \exp \lk - \frac{1}{2}
\int d^3 x F_{i j} (\vek{x}) \frac{1}{\sqrt{-D^2+c}}F_{i j} (\vek{x})\rk \hk ,
\label{10b}
\ee
where $D^2$ is the adjoint covariant Laplace operator. In Ref.~\cite{Greensite:2007ij},
where the $D=2+1$ theory was examined, 
the choice $c = -\lambda_0 + m^2$ was made, $\lambda_0$ being the lowest 
eigenvalue of $-D^2$. The technical difficulties of working with
eq.~(\ref{10b}) are however beyond the scope of this paper, since
one does not expect the Laplacian term to modify the vortex
content of the $D=3$ theory.
Even the simplified version given in
eq.~(\ref{1}) cannot be used for analytic calculations; instead, the
calculation of expectation values in this state requires conventional
3-dimensional Yang-Mills lattice simulations. As stated earlier,  these
$D=3$ lattice calculations must be carried out in Landau gauge when
(\ref{1}) is used as Schr\"odinger wave functional in Coulomb gauge.
\bi

\no
The infrared properties of the vacuum sector of Yang-Mills theory are known
to be dominated by center vortices \cite{R8}. This is also true
in $D = 3$. The wave functional eq.~(\ref{1}) thus contains,
among other configurations, an ensemble of percolating center vortices.
By standard lattice methods \cite{Forcrand:1999ms} we can extract the center
vortex
content of this wave functional. Therefore the use of the wave functional
eq.~(\ref{1}) also allows us to study (on the lattice) how, within the
Hamiltonian approach, the Gribov-Zwanziger confinement mechanism
is related to the center vortex picture of confinement.
Previous lattice calculations \cite{Greensite:2003xf,Greensite:2004ur} have
shown that removal of center vortices from the (4-dimensional)
Yang-Mills ensemble by the method of Ref. \cite{Forcrand:1999ms} makes the ghost
form factor infrared finite, analogously to the suppression observed before in
Landau gauge \cite{Gattnar:2004bf}.
In the present work we will calculate the center vortex contribution to
various Coulomb gauge propagators. The paper is organized as follows:
\bi

\no
In the next section, we briefly recall some properties of $D=3$ Yang--Mills
theory, in particular the scaling behavior and some known facts about
Landau gauge Green functions. Section 3 contains a description of our
numerical setup and the results for the static gluon and ghost form factors
in Coulomb gauge. We also discuss the role of center vortices and their
implications for the Gribov--Zwanziger scenario. We close with a brief summary
and an outlook on future investigations.


\section{Yang-Mills theory in three dimensions}
Since we want to describe the continuum model of eq.~(\ref{1}) using a lattice,
we must first have a closer look at the scaling properties of $D=3$ YM theory.
The lattice model is defined on a $D=3$ cubic space-time grid with periodic
boundary conditions. We will use the Wilson action and employ various gauge fixing
algorithms.
The scale $\mu$ of the continuum wave functional in eq.~(\ref{1}) plays the role of
the three-dimensional bare YM coupling in the continuum limit, $\mu = g_0^2$.

In three-dimensional YM theory, the renormalized coupling $g_R$ is
\[
\beta = \frac{4}{a g_0^2}\,,\qquad\qquad g_0^2 = g_R^2 \left[ 1 +
\mathscr{O}(\hbar \beta^{-1}) \right]
\]
in terms of the bare coupling $g_0$ and the lattice spacing $a$. The string
tension in units of the lattice spacing is therefore
\begin{equation}
\hat{\sigma} \equiv \sigma a^2 = \sigma\,\frac{16}{\beta^2 g_0^4}
= \frac{16\,\sigma}{\beta^2\,g_R^4}\, + \mathscr{O}(\beta^{-3})\,.
\label{23b}
\end{equation}
Since $g_R$ sets the overall scale, $\sigma/g_R^4$ is a dimensionless constant
independent of $\beta$, i.e.~large Creutz ratios should scale according to
\begin{equation}
\chi(R,R) \approx \hat{\sigma}(\beta) = \frac{{\rm const}}{\beta^2} +
\mathscr{O}(\beta^{-3})\,.
\label{20}
\end{equation}
Large-scale simulations in $SU(2)$ \cite{teper:99} have indeed confirmed
the existence of a scaling window $\beta \approx 3\ldots12$ in which the
dependency (\ref{20}) can be observed.
More precisely, the best fit \cite{teper:99} is
\begin{equation}
\hat{\sigma}(\beta) = \frac{1.788}{\beta^2}\,\left(1 +
 \frac{1.414}{\beta} + \cdots \right)\,.
\label{22}
\end{equation}
valid for $\beta \ge 3$. From eqs.~(\ref{23b}), (\ref{22}) and
$\mu = g_0^2$, we find
\begin{equation}
\frac{\sigma}{\mu^2} = 0.111
\,\left[1 + \frac{1.414}{\beta} + \cdots \right]\,,
\label{27}
\end{equation}
i.e.~the variation parameter $\mu = \mu(\sigma,\beta)$ is fixed once
the overall scale $\sigma$ and the lattice coupling $\beta$ in the
scaling window are given. For the standard value
$\sqrt{\sigma} = 440\,\mathrm{MeV}$, $\mu$ falls in the range
$
\mu = 1088\,\mathrm{MeV}\,\,\ldots\,\,1236\,\mathrm{MeV}
$
when $\beta$ is varied in the scaling window.
There is however no compelling reason for the string tension
of this $D=3$ model of the $D=4$ Yang-Mills vacuum to
coincide with the "physical" string tension of the genuine
$D=3$ Yang-Mills theory.
In  view of the following we will then treat $\mu$ as an adjustable
parameter that controls the tradeoff between matching the $D=4$ string
tension or the corresponding Green functions.

Ghost and gluon Green functions in Landau gauge are generally expected to be
multiplicatively renormalizable, i.e.~the lattice data for different $\beta$
must fall on top of each other once the momentum is expressed in physical
units and a finite field renormalization $Z(\beta)$ is applied. The Gribov-Zwanziger
confinement criterion makes qualitative predictions about the deep IR behavior
of Green functions. It is based on the idea that the gauge field
configurations in the functional integral should be restricted to the
so-called fundamental modular region (FMR). The dominant contributions
within the FMR should then come from field configurations near the Gribov
horizon,
where the near-zero modes of the Faddeev-Popov operator\footnote{We write
$\hat{D}^{ab} = \partial_\mu + g\,\hat{A}^{ab} = \partial_\mu - g f^{abc}
A_\mu^c$
for the covariant derivative acting on adjoint color fields.}
$\mathsf{M} \equiv - \partial\hat{D}$ strongly enhance the Coulomb potential
\begin{equation}
V_c(|\vek{x}-\vek{y}|) \sim \mathrm{Tr}\,\left\langle \MM^{-1}\,
(-\Delta)\,\MM^{-1} \right\rangle.
\label{40}
\end{equation}
To enforce the restriction to the FMR, a non-local but renormalizable
horizon term may be added to the Yang-Mills action. The fact that the
partition
function is dominated by near-horizon configurations is then expressed as the so-called
\emph{horizon condition}, which in turn implies that the ghost propagator
should be more singular than the free ghost in the deep infrared,
\begin{eqnarray}
G^{ab}(p) &\equiv& \langle (\mathsf{M}^{-1})^{ab} \rangle =
\frac{\delta^{ab}}{p^2}\,d(p)\nonumber \\[2mm]
\lim\limits_{p \to 0} \,d^{-1}(p) &=& 0\,.
\label{41}
\end{eqnarray}
Similarly, the gluon propagator should be infrared suppressed or vanishing,
\begin{eqnarray}
D^{ab}_{\mu\nu}(p) &\equiv& \langle A_\mu^a(-p)\,A_\nu^b(p) \rangle
= \delta^{ab}\,\left(\delta_{\mu\nu} - \frac{p_\mu \,p_\nu}{p^2}\right)\,
D(p)\nonumber \\[2mm]
\lim\limits_{p\to 0} \,D(p) &<& \infty.
\label{42}
\end{eqnarray}
These results are expected to hold both in $D=3$ and $D=4$ dimensions.
It should be noted, however, that a soft BRS breaking but renormalizable
mass term is possible in $D=3$ and $D=4$ which gives rise to the so-called
\emph{decoupling solution}. The latter is characterized by a tree-level-like
ghost propagator in the IR ($d(0) = \mbox{const}$) accompanied by an infrared
finite,
but non-vanishing gluon propagator $D(0) \neq 0$. Recent lattice investigations
seem to favor this type of behavior in Landau gauge for
both $D=3$ and $D=4$ which is, however, at odds with both the Gribov-Zwanziger
and the Kugo-Ojima criterion favored by functional methods in the continuum.
As for the Coulomb gauge  in $D=4$, there is, however, ample evidence
both from variational methods
\cite{Feuchter:2004gb, Epple:2006hv}
and lattice investigations \cite{Burgio:2008jr,Burgio:2009xp}
that the ghost propagator is infrared enhanced while the static gluon
propagator is
infrared vanishing, in accord with the original Gribov scenario.

In the next section, we compute the gluon and ghost Green functions in the
deep infrared, using the confining wave functional eq.~(\ref{1}) as a model
of the $D=4$ ground state in Coulomb gauge, reliable at least
in some intermediate momentum range. Technically, this amounts to a
lattice simulation in $D=3$ Landau gauge; in contrast to earlier studies
cited above, we will first perform a detour via the maximal center gauge
(MCG) to identify the percolating vortex content of each configuration.
We can then remove or isolate these vortices before going to Landau gauge,
and thus study the interplay between the vortex and Gribov-Zwanziger scenario.


\section{Numerical results}

\subsection{Numerical setup}
Since we are mainly interested in the IR properties we choose in the following
a fixed coupling $\beta = 3.5$ close to the smaller end of the scaling
window, which in turn gives access to the smallest momenta while still
allowing to extract continuum physics. Calculations were carried out on
$20^3$ and $40^3$ lattices using a standard over-relaxed version of Creutz'
heath-bath algorithm. We employed 500 sweeps for thermalization and
50 to 100 sweeps between measurements to reduce auto-correlations. For each
Green function, we took 300 measurements on the smaller volume $20^3$,
and 100 measurements for the volume $40^3$.

In each measurement, the lattice configuration was first brought to
maximal center gauge (MCG) to allow for a clean identification of
the center vortex content. The vortices were then either removed or
isolated, and both fields (in addition to the unmodified MCG configuration)
were subjected to further Landau gauge fixing.\footnote{It should be noted
that MCG corresponds to Landau gauge for the adjoint link, which reduces to
ordinary Landau gauge once the singular vortex background is removed. Thus,
the vortex removed MCG configurations \emph{are} already in Landau gauge.
Nevertheless,
we have carried out the subsequent Landau gauge fixing step to be sure that we
converged to the \emph{best} minimum within our numerical precision.}
As explained above, this corresponds to $D=4$ simulations with the trial wave
functional eq.~(\ref{1}) (including its center vortex content) in Coulomb gauge.

For both gauge fixing steps, we used an iterated over-relaxation
algorithm with up to $30$ restarts after a random gauge transformation
on the initial configuration to reduce the Gribov noise. The gauge fixing
procedure was terminated when the local gauge violation was
sufficiently small,
\[
\|\theta\|_\infty \equiv
\max_x \max_{k \in \{1,2,3\}} \left|\theta^k(x)\right|
< \epsilon = 10^{-12}\,,
\]
where in terms of the the quaternion representation $U_\mu = a_\mu^0 + i
a_\mu^k\,\sigma_k$ of the $SU(2)$ links,
\[
\begin{array}{l@{\qquad}r@{\,\,\,=\,\,\,}l}
\mathrm{MCG: }
& \displaystyle \theta^k(x) &
\displaystyle \frac{1}{2}\sum_{\mu=0}^2\left[ a_\mu^0(x)\,
a_\mu^k(x) - a_\mu^0(x-\hat{\mu})\,a_\mu^k(x-\hat{\mu}) \right]
\\[6mm]
\mathrm{Landau: }
& \displaystyle \theta^k(x) &
\displaystyle \frac{1}{2} \sum_{\mu=0}^{2} \left( a_\mu^k(x) -
a_\mu^k(x-\hat{\mu}) \right)\,.
\end{array}
\]
In all cases, the gauge fixing functional was numerically stationary
long before the violation limit was reached.

For Landau gauge, we further improved the quality of the gauge
fixing using either flip pre-conditioning steps \cite{precondition} and, on the
larger
$40^3$ lattice, simulated annealing methods. However, these improvements
had only a marginal effect on the attained maximum of the gauge fixing functional,
since the previous MCG fixing acts as a strong pre-conditioning even when
vortices were \emph{not} removed.

\subsection{Gluon propagator}
The gauge field on the lattice is extracted using the standard definition,
\[
a(\beta)\,A^k_\mu(x + \hat{\mu}/2) \approx \frac{1}{2i}\,\left(U_\mu(x) -
U_\mu^\dagger(x)\right) = 2 a_\mu^k(x)
\]
where $a(\beta)$ is the lattice spacing and $U_\mu = a_\mu^0 + i\,a_\mu^k\,
\sigma_k$
is the quaternion representation of the links.\footnote{This assumes that the
gauge fields are sufficiently smooth and the links
close to unity, $a_\mu^0 \approx 1$, as ensured by Landau gauge and the
continuum limit.} A fast Fourier transformation
then gives access to the gluon propagator eq.~(\ref{42}).\footnote{Since only a
single coupling $\beta=3.5$ was considered, no further renormalization was
necessary. On the smaller lattice, we have verified that all considered
ensembles do lead to multiplicatively renormalizable gluon and ghost
propagators.}

\begin{figure}[t]
\vspace*{3mm}
\begin{center}
\includegraphics[width=7.5cm]{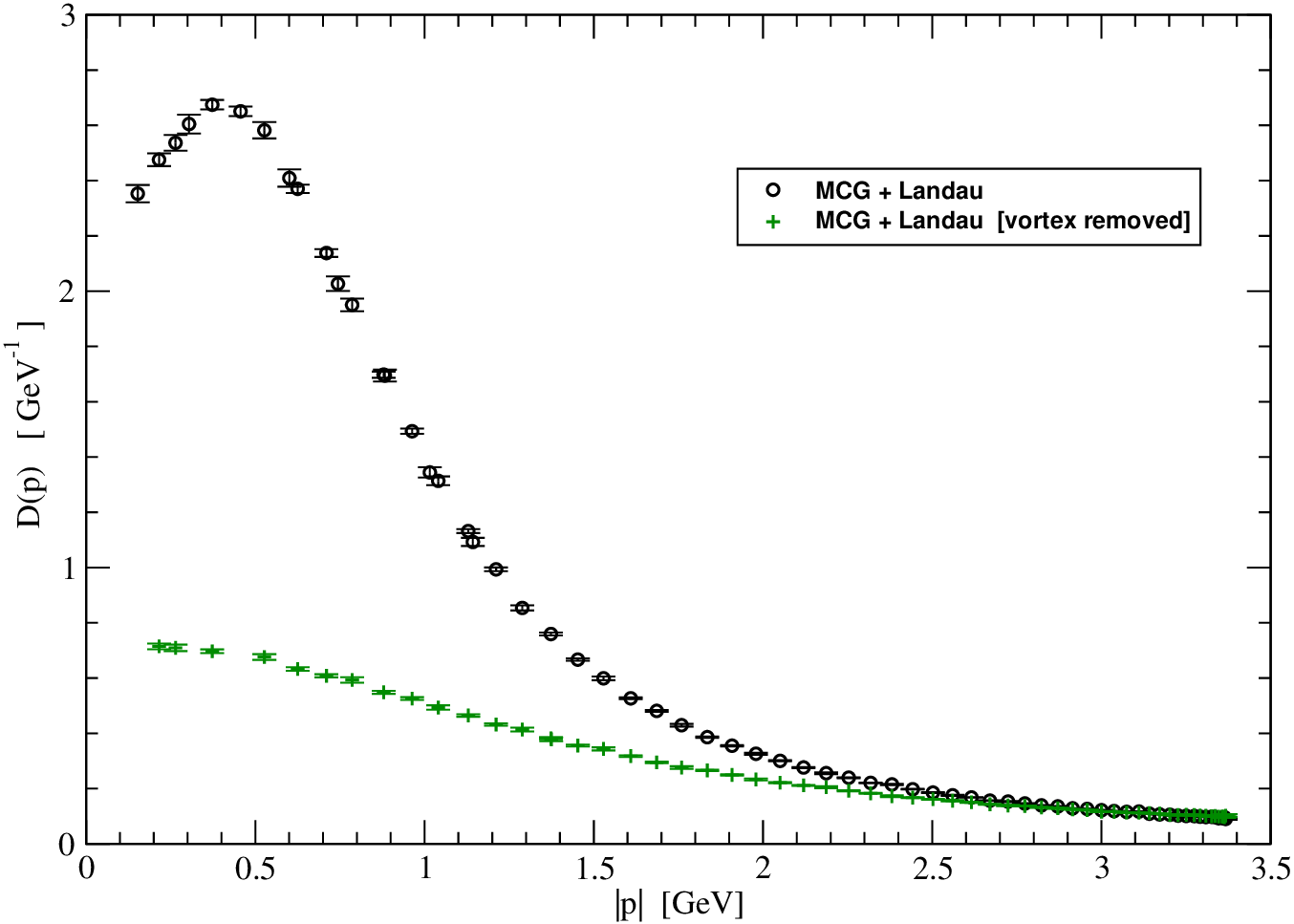}
\hfill
\includegraphics[width=7.5cm]{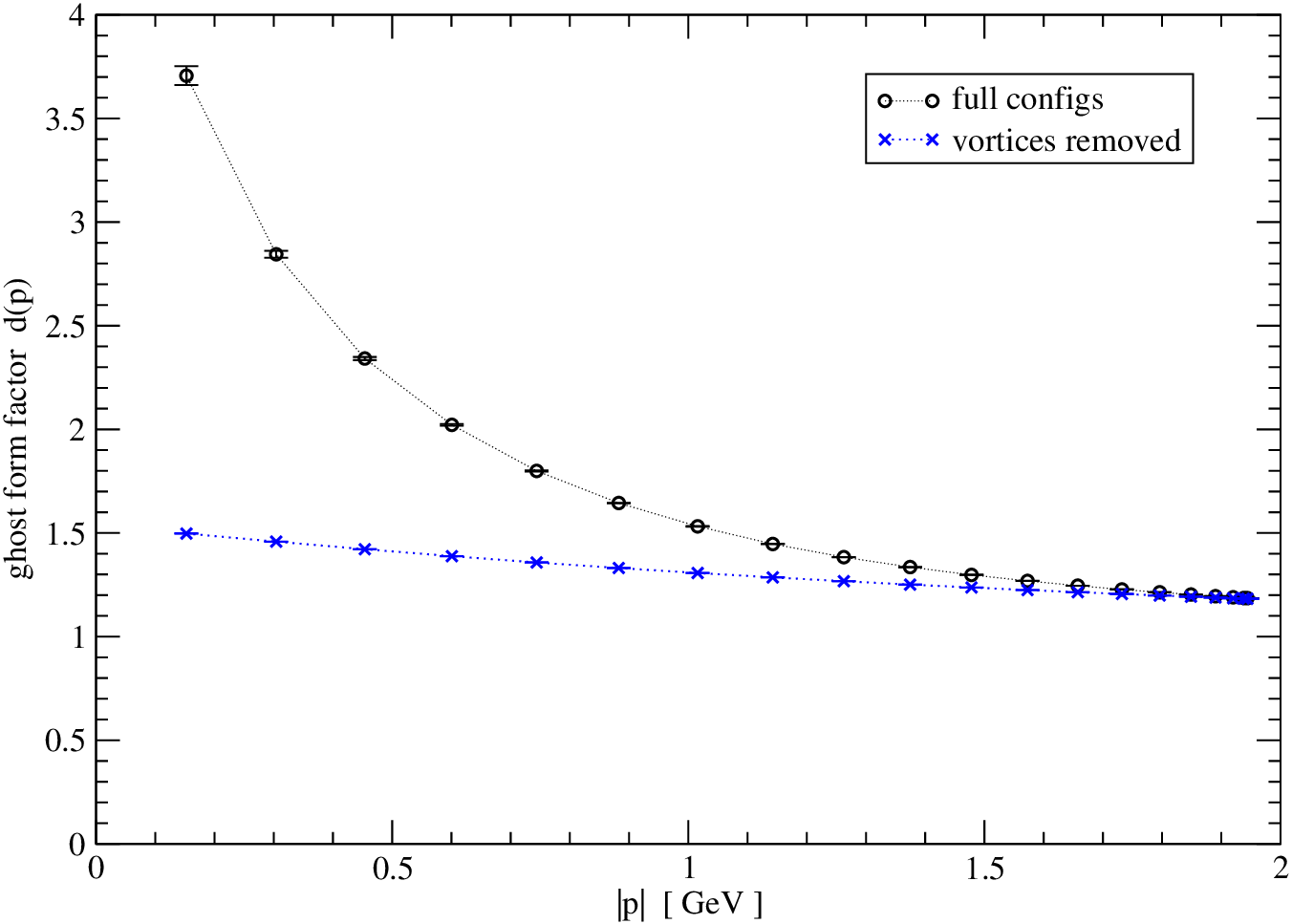}
\end{center}
\caption{\label{fig:1}\small Left panel: The $D=3$ gluon propagator in Landau
gauge after prior
MCG fixing, both with and without vortices. Right panel: The same plot for
the $D=3$ ghost form factor in Landau gauge.}
\end{figure}

In the left panel of Fig.~\ref{fig:1}, we first show the $D=3$ gluon
propagator
in Landau gauge, after prior MCG fixing. As long as vortices are not removed,
the resulting propagator is virtually identical with the one obtained by
fixing minimal Landau gauge directly. As can be clearly seen, the removal
of vortices has little effect in the UV but clearly suppresses the propagator
at intermediate and small momenta.

\begin{figure}[t]
\vspace*{3mm}
\begin{center}
\includegraphics[width=7.5cm]{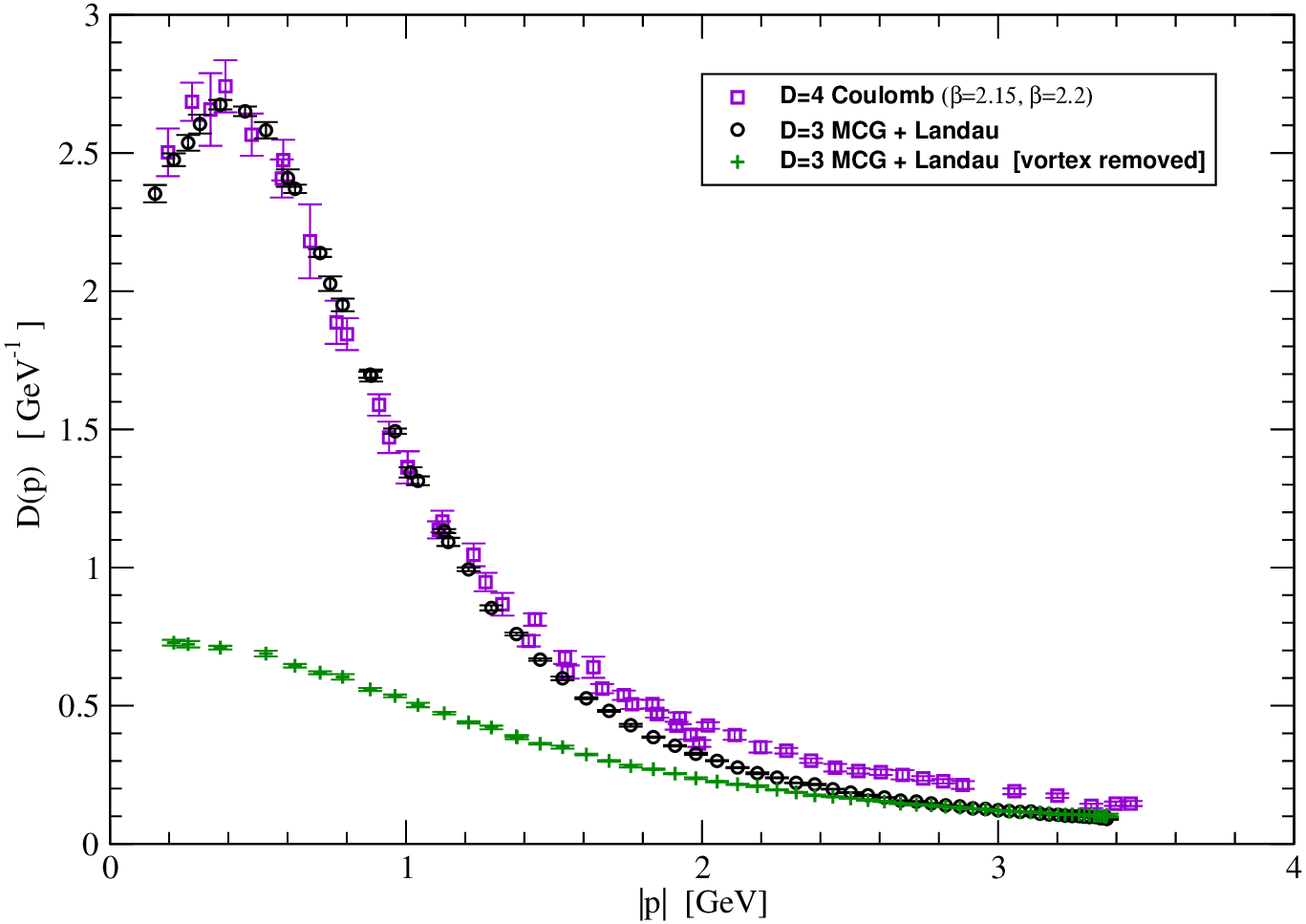}
\hfill
\includegraphics[width=7.5cm]{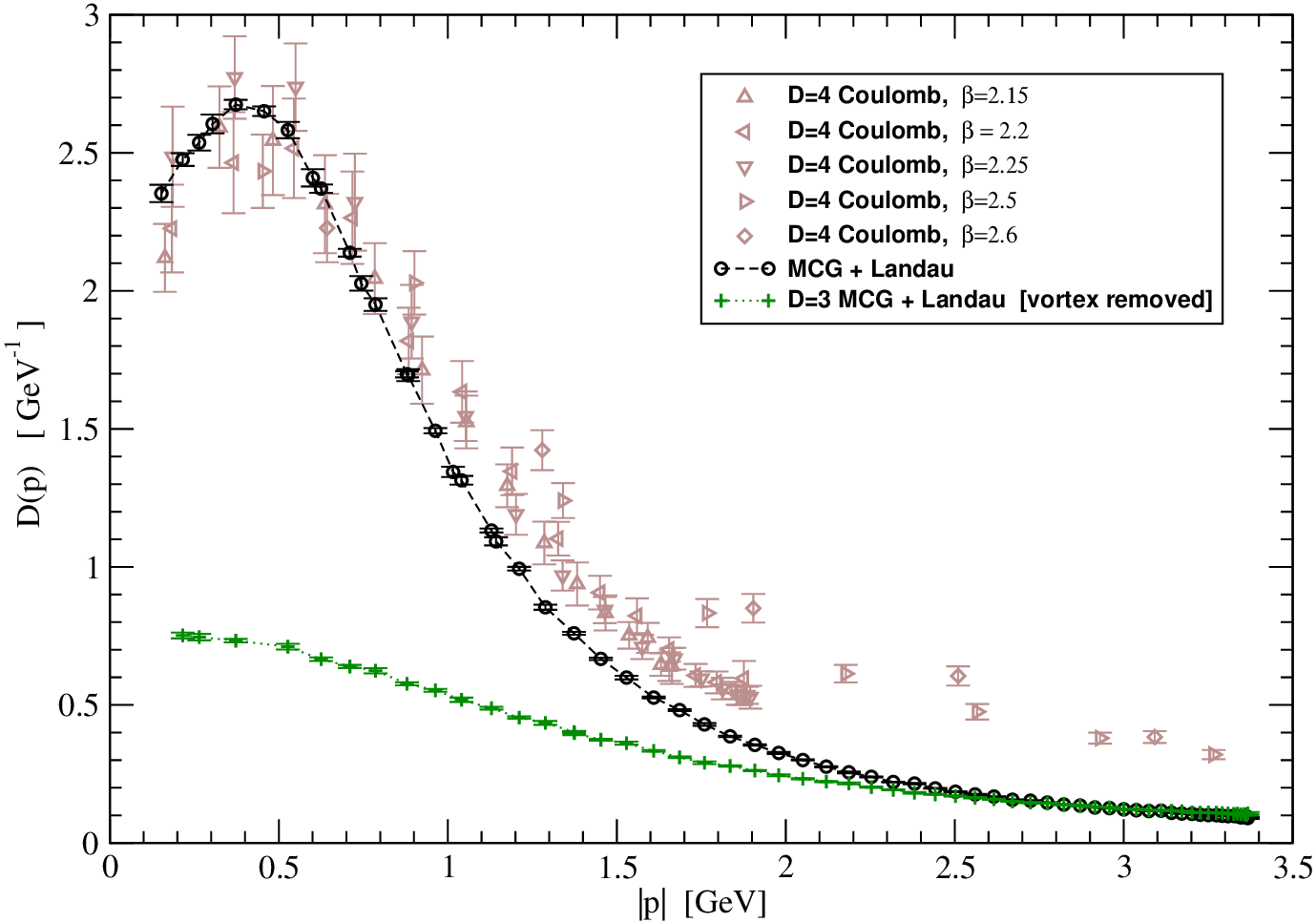}
\end{center}
\caption{\label{fig:2}\small The equal-time gluon propagator in $D=4$ Coulomb
gauge, compared to the $D=3$ gluon propagator in MCG+Landau gauge,
both with and without vortices. In the left panel, we have only included
$D=4$ data from two close couplings, which hides the inherent scaling
violations. If more couplings are included (right panel), the scaling
problems become apparent and the comparison with the $D=3$ data is less
favorable.}
\end{figure}

Fig.~\ref{fig:2} shows the equal-time gluon propagator
$D_C(\vek{p})$ for Coulomb gauge in $D=4$, as obtained \emph{before} removal
of scaling violations.  We have compared this data to the  $D=3$ gluon
propagator
\begin{enumerate}
\item[\textbf{a.}] directly in (minimal) Landau gauge
\item[\textbf{b.}] in Landau gauge after prior MCG fixing,
\item[\textbf{c.}] in Landau gauge after prior MCG fixing and vortex removal
\end{enumerate}
(The cases \textbf{a.} and \textbf{b.} are virtually identical, and only
\textbf{b.} is shown in Fig.~\ref{fig:2} to make the plot less cluttered.)
The confining $D=3$ MCG/Landau propagators  $D_L(p)$ \textbf{a.}  and
\textbf{b.} show good agreement in the low and intermediate momentum range
with the \emph{naive} instantaneous $D=4$ Coulomb propagator,
once the (Wilson) string tensions in the two calculations are set
equal.\footnote{In the UV, we have the expected deviations, since the $1/p$
behavior of the Coulomb result decays much slower than the $1/p^2$ typical
for the Landau case.} (We took $\sqrt{\sigma} = 440\,\mathrm{MeV}$ to set the
scale on the momentum  axis.) As expected, the non-confining Landau gauge
propagator with  vortices removed \emph{cannot} be matched with
the $D=4$ Coulomb data in \emph{any} momentum range.

However, this agreement of the confining propagators must be considered
spurious,
since the naive $D=4$ Coulomb data is subject to severe scaling violations.
In Ref.~\cite{Burgio:2008jr}, these violations were thoroughly analyzed and a
procedure to extract the true instantaneous propagator $D_C(\vek{p})$ in the
infinite volume and continuum limit was devised. The result displays proper
scaling and the correct deep ultra-violet behavior
$D_C(\vek{p})\sim |\vek{p}|^{-1}$; it can also be fitted in the entire momentum
range through the Gribov formula \cite{Burgio:2008jr}.

\begin{figure}[t]
\vspace*{3mm}
\begin{center}
\includegraphics[width=7.5cm]{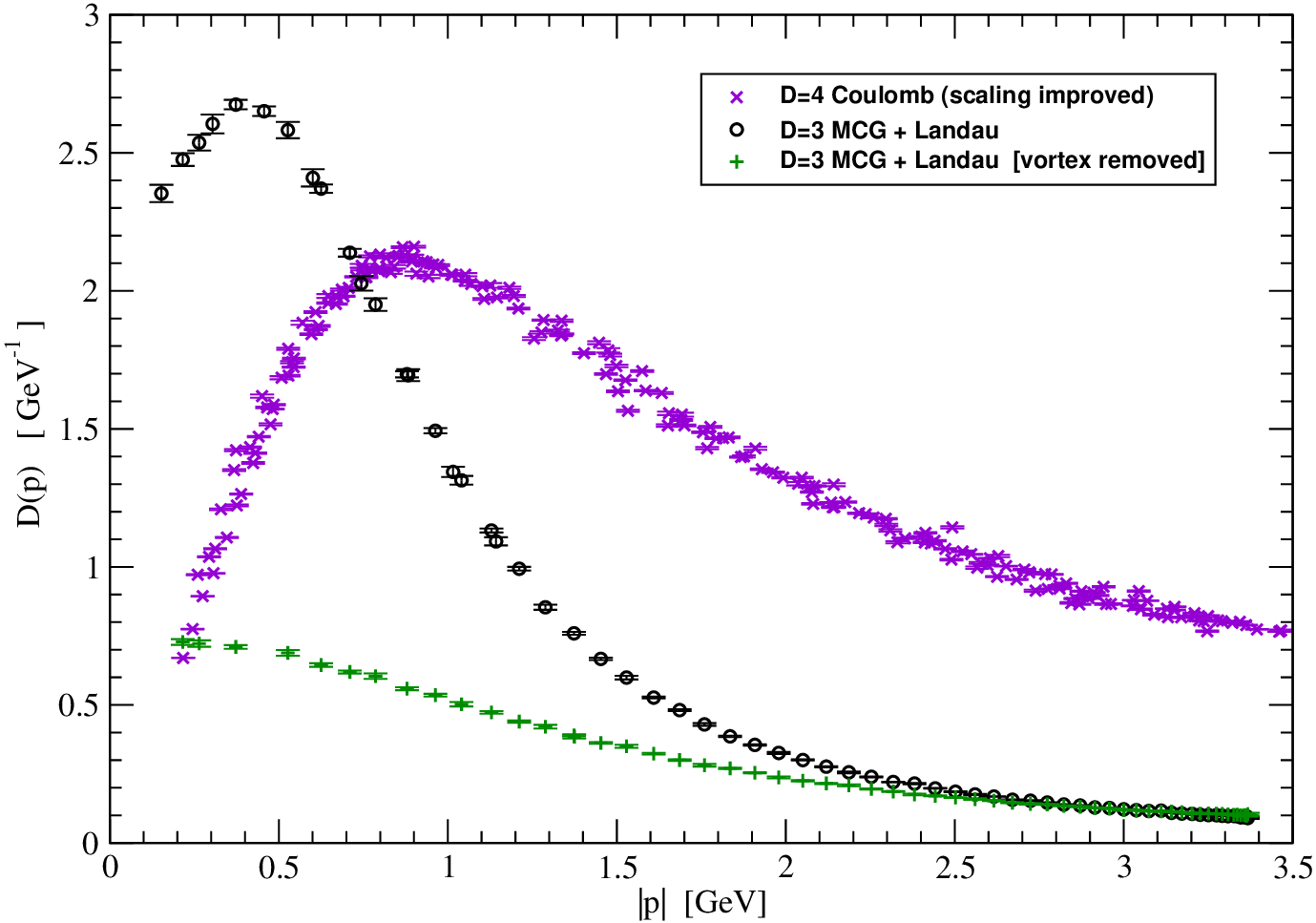}
\hfill
\includegraphics[width=7.5cm]{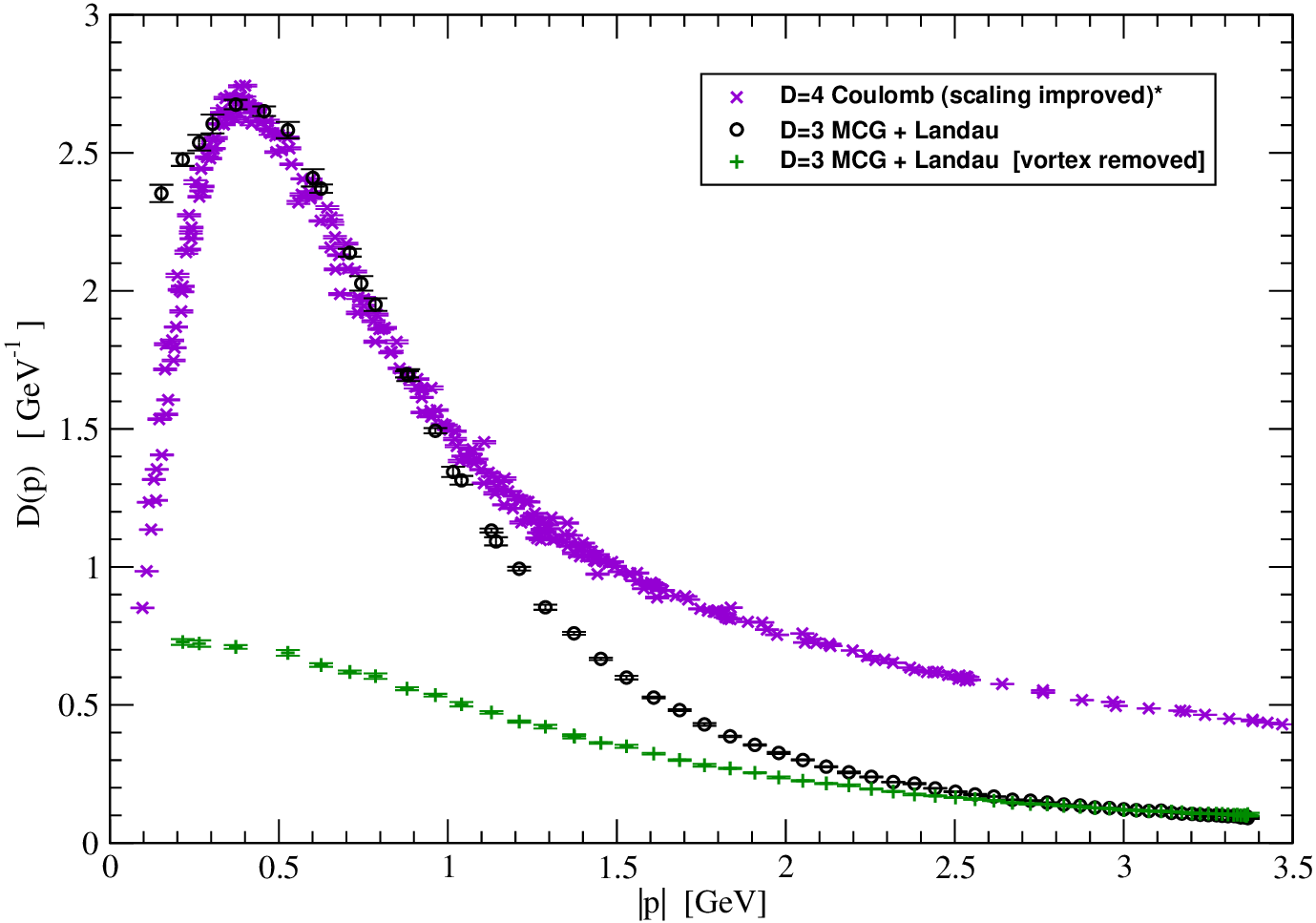}
\end{center}
\caption{\label{fig:3}\small The equal-time gluon propagator in $D=4$ Coulomb
gauge, compared to the $D=3$ gluon propagator in MCG+Landau gauge,
both with and without vortices. The scaling violations in the
Coulomb data are removed \cite{Burgio:2008jr} to exhibit the true
continuum result. In the left panel, the (Wilson) string tension of the
$D=3$ and $D=4$ calculations are matched, while the right panel has
$\sigma_{3D} \approx 6.2\,\sigma_{4D}$. (For simplicity, we have
scaled $\sigma_{4D}$ and left $\sigma_{3D}$ at the standard value.)}
\end{figure}

As can be seen from the left panel in Fig.~\ref{fig:3}, the improved
static Coulomb propagator $D_C(\vek{p})$ differs considerably from the naive
result,
so that the spurious agreement with the confining $D=3$ ensembles is destroyed.
In particular, the deviation between the strong $1/p^2$ decay of the $D=3$
propagator and the slow $1/|\vek{p}|$ decay of $D_C(\vek{p})$ is now much more
apparent. Qualitative agreement in the phenomenologically important momentum
range around $1\,\mathrm{GeV}$ can only be reached if the $D=3$ momentum is
scaled with a factor $\sim 2.5$ as compared to the $D=4$ case (see right panel
in fig.~\ref{fig:3}). This in turn would mean that the scale parameter $\mu$
in the initial wave functional eq.~(\ref{1}) must be altered by
the same factor and the $D=3$ string tension would no longer match the $D=4$
result.\footnote{In the very deep IR, the $D=3$ and $D=4$ propagators cannot
be matched even with the relaxed condition on $\mu$, since the
correct $D=4$ static Coulomb propagator vanishes as $|\vek{p}|\to 0$, while the
$D=3$ MCG-Landau propagator $D_L(p)$ goes to a non-zero value in this limit.}
The best agreement is thus obtained if $\sigma_{3D} \approx 6.2\,\sigma_{4D}$,
corresponding to $\mu \approx 7\,\mathrm{GeV}$.

The vortex-removed ensemble \textbf{d.} leads to a gluon propagator which
is incompatible with the $D=4$ Coulomb result in \emph{any} momentum range,
even if the restrictions on the parameter $\mu$ are relaxed. This shows that
the percolating center vortex content is not only indispensable for the
confining properties of the Wilson loop, but also for the qualitative
behavior of the gluon propagator at intermediate momenta.

As stated above, a better approximation of the correct vacuum wave functional
is given
by eq.~(\ref{10b}). Indeed, either by inspecting Fig.~\ref{fig:3} or on simple
dimensional arguments, it should be clear that one should rather compare 
the $D=4$ static Coulomb propagator
$D_C(\vek{p})$ with $p\, D_L(p)$ \cite{Burgio:2009xp}, which on the other hand will 
mimic the somewhat less crude approximation of the vacuum wave functional:
\be
\Psi [\vek{A}] = \cN \exp \lk - \frac{1}{2}
\int d^3 x F_{i j} (\vek{x}) \frac{1}{\sqrt{-\nabla^2}}F_{i j} (\vek{x})\rk
\hk ,
\label{10c}
\ee
\begin{figure}[t]
\vspace*{3mm}
\begin{center}
\includegraphics[width=14cm]{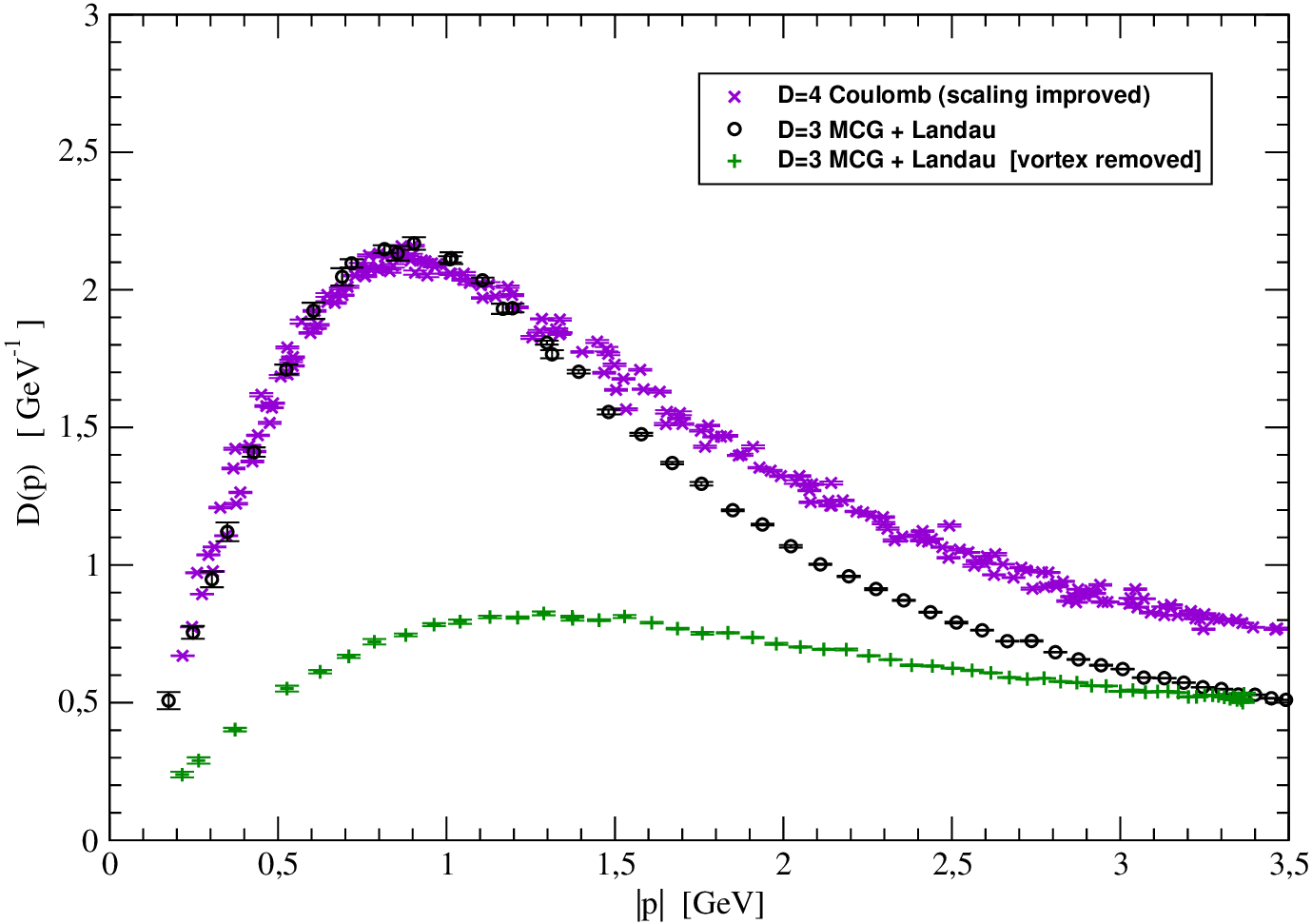}
\end{center}
\caption{\label{fig:5}\small The gluon static propagator in $D=4$ Coulomb
gauge,
compared to the approximation $p\, D_L(p)$ of eq.~(\ref{10c})
in $D=3$ MCG+Landau gauge.}
\end{figure}
The data basically coincide when tuning $\sigma_{3D} \approx 1.49\,\sigma_{4D}$,
as can be seen in fig.~\ref{fig:5}. Such good agreement in the
direction
of the wave functional given in eq.~(\ref{10b}) definitely deserves further
investigation.

\subsection{The ghost form factor}

Although
large volume simulations performed in a setup similar to ours 
clearly show a decoupling behavior also for the Landau
ghost in $D=3$
\cite{decoupling}, for the
momenta available in our simulations this regime still has not kicked in,
so that eq.~(\ref{1}) might still be considered a valid approximation in the
following.
Future simulations closer to the thermodynamic/continuum limit
will however explicitely have to deal
with eq.~(\ref{10b}) or at least eq.~(\ref{10c}), since the simple
dimensional argument leading for the static gluon to the comparison in
fig.~\ref{fig:5} cannot obviously be applied to the Faddeev-Popov operator.

\begin{figure}[t]
\vspace*{3mm}
\begin{center}
\includegraphics[width=14cm]{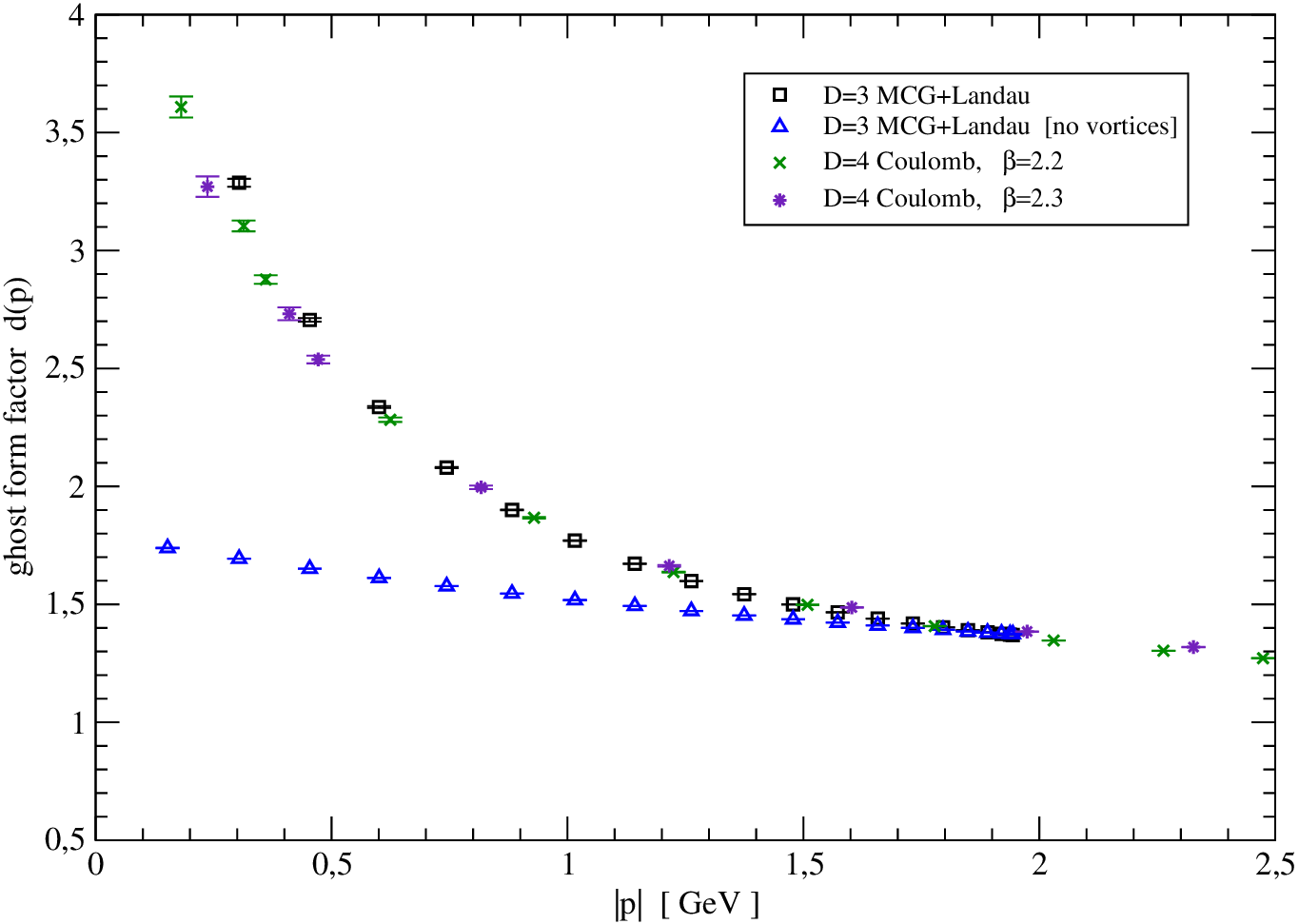}
\end{center}
\caption{\label{fig:4}\small The ghost form factor in $D=4$ Coulomb gauge,
compared to the
same quantity in $D=3$ MCG+Landau gauge. The $D=4$ data is from a $32^4$
lattice
at $\beta=2.2$ and $\beta=2.3$, while the $D=3$ data was obtained on $40^3$
lattices with $\beta=3.5$.}
\end{figure}

The right panel of figure \ref{fig:1} shows the dimensionless ghost form factor
(\ref{41}) as obtained from $D=3$ Landau gauge with prior MCG fixing.
Without vortex removal, the result is virtually identical to the direct
Landau gauge fixing, and a power law behavior of the form
\begin{equation}
d(p) \sim p^{-2 \kappa}\,\qquad\qquad \kappa = 0.23(1)
\label{conf}
\end{equation}
can be fitted. This agrees well with the infrared exponent of
$D=4$ Coulomb gauge \cite{Quandt:2007qd}, which is not subject to scaling
violations. Fig.~\ref{fig:4} compares the ghost form factors in $D=3$
Landau gauge and $D=4$ Coulomb gauge quantitatively; as expected from the
matching IR exponent, the comparison is fairly good.
In the UV, the
$d(p)$ approaches $1$ in both $D=3$ and $D=4$, so the qualitative agreement
is also good in this regime. (A closer comparison of possible anomalous
dimensions in the sub-leading terms is beyond our current numerical precision.)

It should be noted that the power-like behavior eq.~(\ref{conf}) is
conformally invariant, i.e.~the exponent $\kappa$ is not affected by any
rescaling of the momenta or string tension. This allows to maintain the good
agreement of the ghost form factor, while relaxing the condition on $\mu$ to
make the gluon propagators match (approximately). With the higher value
of $\mu$ and the $3D$ string tension, it is therefore possible to reproduce
at least qualitatively both the gluon and ghost form factor from the
wave functional eq.~(\ref{1}), at least in the intermediate momentum range.

From the right panel in Fig.~\ref{fig:1}, it is also seen that the
vortex-removed
ghost form factor in $D=3$ is virtually flat ($\kappa < 0.05$) so that it
cannot be matched even qualitatively to the $D=4$ Coulomb result.
In particular, it seems impossible to satisfy the horizon condition
$d(0)^{-1} = 0$ once center vortices are removed.
This is, of course, expected since vortices live on the Gribov horizon and are
thus indispensable to maintain enough near-zero modes of the Faddeev-Popov
operator to produce the diverging ghost form factor. A similar conclusion was
also reached in Ref.~\cite{Greensite:2004ur} from the direct study of the
low-lying Faddeev-Popov spectrum.


\section{Conclusion and outlook}

In this paper we have investigated the properties of the
static propagators of the Hamiltonian approach in Coulomb gauge, assuming the
wave functional eq.~(\ref{1}), which shows strict confinement, as
an approximation of the low energy limit of the true Yang-Mills vacuum wave
functional.
The calculation of expectation values of observables in the
$D=3+1$ Hamiltonian approach in Coulomb gauge in this state
requires and ordinary $D=3$ lattice simulation in Landau gauge. Adjusting the
free scale $\mu$ of the state eq.~(\ref{1}) yields propagators which in a low to
intermediate momentum regime (up to about $1~\mathrm{GeV}$) reproduce
quite well the exact lattice propagators of the $D=3+1$ Coulomb gauge,
which also agree qualitatively with propagators obtained in the variational
approach to continuum Yang-Mills theory in Coulomb gauge. An even better agreement
should be obtained by employing the wave functional given in eq.~(\ref{10b}), (\ref{10c}),
as Fig.~\ref{fig:5} shows.

The propagators drastically change in the infra-red when the magnetic center
vortices occurring with the weight $|\Psi[\vek{A}]|^2$ in the wave functional
are removed by the method of Ref.~\cite{Forcrand:1999ms}. In
particular,
the  ghost form factor loses its infrared singularity and the horizon condition
is no longer satisfied. These results indicate that there is no need to
include additional explicit center vortex degrees of freedom in the trial wave
wave functional of the variational approach, as was proposed recently
\cite{sczcepaniak}.

\section*{Acknowledgments} 
We would like to thank Jeff Greensite for 
careful reading of the manuscript and useful 
suggestions. 
This work was partly supported by DFG under 
the contract DFG-Re856/6-3. 



\begin{thebibliography}{99}

  \bibitem{R8}
  For lattice evidence for the center vortex picture of confinement see: L.
  DelDebbio, M. Faber, J. Greensite and S. Olejnik, Phys. Ref. {\bf D55}, 2298
  (1997); K. Langfeld, H. Reinhardt and O. Tennert, Phys. Lett. {\bf B419},
  316,
  (1998); M. Engelhardt, K. Langfeld, H. Reinhardt and  O. Tennert, Phys. Rev.
  {\bf D61}, 054504 (2000); J. Greensite, Prog. Part. Nucl. Phys. {\bf 51}, 1
  (2003)

   \bibitem{RX1}
For lattice evidence for the dual super-conductor picture of confinement see
A.S. Kronfeld, G. Schierholz and U.-J. Wiese, Nucl. Phys. {\bf B198} (1987)
516;
T. Suzuki, I. Yotsuyanagi, Phys. Ref. {\bf D42} (1990) 4257; S. Hoki et al.
Phys. Lett. {\bf B272} (1991) 326; G. Bali, Ch. Schlichter, K. Schilling, Prog.
Theor. Phys. Suppl. {\bf 131} (1998) 645 and Ref. therein.

\bibitem{Nambu:1974zg}
  Y.~Nambu,
  Phys.\ Rev.\  D {\bf 10}, 4262 (1974);
  S.~Mandelstam,
  Phys.\ Lett.\  B {\bf 53}, 476 (1975);
  G.~Parisi,
  Phys.\ Rev.\  D {\bf 11}, 970 (1975);
  Z.~F.~Ezawa and H.~C.~Tze,
  Nucl.\ Phys.\  B {\bf 100}, 1 (1975);
  R.~Brout, F.~Englert and W.~Fischler,
  Phys.\ Rev.\ Lett.\  {\bf 36}, 649 (1976);
  F.~Englert and P.~Windey,
  Nucl.\ Phys.\  B {\bf 135}, 529 (1978);
  G.~'t Hooft,
  Nucl.\ Phys.\  B {\bf 190}, 455 (1981), Phys. Scr. {\bf 25} (1982) 133



\bibitem{'t:1977hy}
  G.~'t ,
  Nucl.\ Phys.\  B {\bf 138}, 1 (1978);
  G.~Mack and V.~B.~Petkova,
  Annals Phys.\  {\bf 123}, 442 (1979);
  G.~Mack,
  Phys.\ Rev.\ Lett.\  {\bf 45}, 1378 (1980);
  G.~Mack and V.~B.~Petkova,
  Annals Phys.\  {\bf 125}, 117 (1980);
  G.~Mack,
  G.~Mack and E.~Pietarinen,
  Nucl.\ Phys.\  B {\bf 205}, 141 (1982);
  Y.~Aharonov, A.~Casher and S.~Yankielowicz,
  Nucl.\ Phys.\  B {\bf 146}, 256 (1978);
  J.~M.~Cornwall,
  Nucl.\ Phys.\  B {\bf 157}, 392 (1979);
  H.~B.~Nielsen and P.~Olesen,
  Nucl.\ Phys.\  B {\bf 160}, 380 (1979);
  J.~Ambjorn and P.~Olesen,
  Nucl.\ Phys.\  B {\bf 170}, 60 (1980);
    E.T. Tomboulis, Physi. Rev. {\bf D 23}, 2371 (1981)

   \bibitem{Greensite:2003xf}
     J.~Greensite and S.~Olejnik, Phys. Rev. {\bf D67}, 094503 (2003)
     [arXiv:hep-lat/0302018].

  \bibitem{Greensite:2004ur}
  J.~Greensite, S.~Olejnik and D.~Zwanziger,
  JHEP {\bf 0505}, 070 (2005)
  [arXiv:hep-lat/0407032].

\bibitem{Reinhardt:2008ek}
  H.~Reinhardt,
  Phys.\ Rev.\ Lett.\  {\bf 101}, 061602 (2008)
  [arXiv:0803.0504 [hep-th]].

  \bibitem{Zwanziger:2002sh}
  D.~Zwanziger,
  Phys.\ Rev.\ Lett.\  {\bf 90}, 102001 (2003)
  [arXiv:hep-lat/0209105].


  \bibitem{Langfeld:2004qs}
  K.~Langfeld and L.~Moyaerts,
  Phys.~Rev.~\textbf{D70} (2004) 074507
  [arXiv:hep-lat/0406024].

  \bibitem{Quandt:2007qd}
  M.~Quandt, G.~Burgio, S.~Chimchinda and H.~Reinhardt,
  PoS {\bf LAT2007}, 325 (2007)
  [arXiv:0710.0549 [hep-lat]].

  \bibitem{Voigt:2007wd}
  A.~Voigt, M.~Ilgenfritz, M.~M{\"u}ller-Preussker, A.~Sternbeck,
  PoS \textbf{LAT2007}, 338 (2007)
  [arXiv:0709.4585]

  \bibitem{Voigt:2008rr}
  A.~Voigt, M.~Ilgenfritz, M.~M{\"u}ller-Preussker, A.~Sternbeck,
  Phys.~Rev.~\textbf{D78} (2008) 014501.
  [arXiv:arXiv:0803.2307 [hep-lat]].

  \bibitem{Burgio:2008jr}
  G.~Burgio, M.~Quandt and H.~Reinhardt, Phys.~Rev.~Lett.~\textbf{102}
  (2009) 032002.

  \bibitem{Burgio:2009xp}
  G.~Burgio, M.~Quandt and H.~Reinhardt, arXiv:0911.5101 [hep-lat] (2009)\,.

  \bibitem{Szczepaniak:2001rg}
  A.~P.~Szczepaniak and E.~S.~Swanson,
  Phys.\ Rev.\  D {\bf 65}, 025012 (2002)
  [arXiv:hep-ph/0107078].

  \bibitem{Feuchter:2004mk}
  C.~Feuchter and H.~Reinhardt,
  Phys.\ Rev.\  D {\bf 70}, 105021 (2004)
  [arXiv:hep-th/0408236].

  \bibitem{Feuchter:2004gb}
  C.~Feuchter and H.~Reinhardt,
  arXiv:hep-th/0402106.

  \bibitem{Reinhardt:2004mm}
  H.~Reinhardt and C.~Feuchter,
  Phys.\ Rev.\  D {\bf 71}, 105002 (2005)
  [arXiv:hep-th/0408237].

  \bibitem{Epple:2006hv}
  D.~Epple, H.~Reinhardt and W.~Schleifenbaum,
  Phys.\ Rev.\  D {\bf 75}, 045011 (2007)
  [arXiv:hep-th/0612241].

  \bibitem{Epple:2007ut}
  D.~Epple, H.~Reinhardt, W.~Schleifenbaum and A.~P.~Szczepaniak,
  Phys.\ Rev.\  D {\bf 77}, 085007 (2008)
  [arXiv:0712.3694 [hep-th]].

  \bibitem{Reinhardt:2008ij}
  H.~Reinhardt and W.~Schleifenbaum,
  Annals Phys.\  {\bf 324}, 735 (2009)
  [arXiv:0809.1764 [hep-th]].

  \bibitem{Schleifenbaum:2006bq}
  W.~Schleifenbaum, M.~Leder and H.~Reinhardt,
  Phys.\ Rev.\  D {\bf 73}, 125019 (2006)
  [arXiv:hep-th/0605115].

  \bibitem{Reinhardt:2007wh}
  H.~Reinhardt and D.~Epple,
  Phys.~Rev.~\textbf{D76} (2007) 065015,
  [arXiv:0706.0175 [hep-th]]\,.

  \bibitem{Feuchter:2007mq}
  C.~Feuchter and H.~Reinhardt,
  Phys.\ Rev.\  D {\bf 77}, 085023 (2008),
  [arXiv:0711.2452 [hep-th]].

  \bibitem{R19}
  M. Leder, J. Pawlowski, A.~Weber and H. Reinhardt, to be published

  \bibitem{Pak:2009em}
  M.~Pak and H.~Reinhardt, arXiv:0910.2916 [hep-th] (2009)\,.

  \bibitem{R20}
  J.~Greensite,
  Nucl.~Phys.~\textbf{B158} (1979) 469.

  \bibitem{decoupling}
  A.~Cucchieri and T.~Mendes,
  Phys.~Rev.~\textbf{D78} (2008) 094503  [arXiv:0804.2371] ; \\
  A.~Cucchieri and T.~Mendes,
  Phys.~Rev.~Lett.~\textbf{100} (2008) 241601  [arXiv:0712.3517] ; \\
  I.L.~Bogolubsky et al.,
  PoS, \textsl{LATTICE2009}, (2009) 237  [arXiv:0912.2249].

  \bibitem{Greensite:2007ij}
  J.~Greensite and S.~Olejnik,
  Phys.~Rev.~D {\bf 77}, 065003 (2008),
  [arXiv:0707.2860 [hep-lat]].

  \bibitem{Forcrand:1999ms}
  P.~de Forcrand and M.~D'Elia,
  Phys.~Rev.~Lett.~\textbf{82} (1999) 4582.

 \bibitem{Gattnar:2004bf}
  J.~Gattnar, K.~Langfeld and H.~Reinhardt,
  Phys.\ Rev.\ Lett.\  {\bf 93}, 061601 (2004)
  [arXiv:hep-lat/0403011]\\
  K.~Langfeld, G.~Schulze and H.~Reinhardt,
  Phys.\ Rev.\ Lett.\  {\bf 95}, 221601 (2005)
  [arXiv:hep-lat/0508007].

  \bibitem{teper:99}
  M.J.~Teper, Phys.~Rev.~\textbf{D59} (1999) 014512.

  \bibitem{precondition}
  I.L.~Bogolubsky et al.,
  Phys.~Rev.~\textbf{D74} (2006) 034503
  [arXiv:hep-lat/0511056] ; \\
  I.L.~Bogolubsky et al.,
  Phys.~Rev.~\textbf{D77} (2008) 014504
  [arXiv:0707.3611]

  \bibitem{sczcepaniak}
  A.~Sczcepaniak, talk given at the ECT workshop
  \textsl{QCD Green's functions, confinement and phenomenology},
  Trento, September 07--11, 2009\,.

  \bibitem{scaling}
  L.~v.~Smekal, R.~Alkofer and A.~Hauck,
  Phys.~Rev.~Lett.~\textbf{79} (1997) 3591  [arXiv:hep-ph/9705242]; \\
  C.~Fischer and J.~Pawlowski,
  Phys.~Rev.~\textbf{D75} (2007) 025012  [arXiv:hep-th/0609009];\\
  C.~Fischer, A.~Maas, J.~Pawlowski,
  Ann.~Phys.~\textbf{324} (2009)  [arXiv:0810.1987].


  \end{thebibliography}
\end{document}